\documentstyle[11pt,fleqn]{article}
\def\titolo{\par\bigskip\begin{center}\bf\LARGE}
\def\endtitolo{\end{center}\par\bigskip\par\rm\normalsize}
\def\instit{\begin{center}\large}
\def\endinstit{\end{center}\rm\normalsize}
\def\references{\begin{thebibliography}{10}}
\def\endreferences{\end{thebibliography}\end{document}}

\newcommand{\btit}{\begin{titolo}}
\newcommand{\etit}{\end{titolo}}

\renewcommand{\author}[1]{\begin{center}\Large #1\end{center}}
\renewcommand{\date}[1]{\par\bigskip\par\sl\hfill #1\par\medskip\par}

\newcommand{\pacs}[1]{\smallskip\noindent{\sl PACS number(s):
                       \hspace{0.3cm}#1}\par\bigskip}
\newcommand{\babs}{\hrule\par\begin{description}\item{Abstract: }\it}
\newcommand{\eabs}{\par\end{description}\hrule\par\medskip\rm}
\newcommand{\ack}[1]{\par\section*{Acknowledgments} #1}
\newcommand{\hs}{\qquad\qquad}         
\newcommand{\nn}{\nonumber}            
\newcommand{\beq}{\begin{eqnarray}}    
\newcommand{\eeq}{\end{eqnarray}}      
\newcommand{\beqn}{\begin{eqnarray}}   
\newcommand{\eeqn}{\end{eqnarray}}     
\newcommand{\at}{\left(}               
\newcommand{\aq}{\left[}               
\newcommand{\ag}{\left\{}              
\newcommand{\ct}{\right)}              
\newcommand{\cq}{\right]}              
\newcommand{\cg}{\right\}}             
\newcommand{\ii}{\infty}                         
\newcommand{\fr}[2]{\mbox{$\frac{#1}{#2}$}}      
\newcommand{\Tr}{\,\mbox{Tr}\,}                  
\renewcommand{\Re}{\,\mbox{Re}\,}                
\newcommand{\al}{\alpha}
\newcommand{\be}{\beta}

\newcommand{\de}{\delta}
\newcommand{\ep}{\varepsilon}
\newcommand{\ze}{\zeta}

\newcommand{\si}{\sigma}

\newcommand{\Ga}{\Gamma}

\begin{document}

\begin{center}

{\large \bf ASYMPTOTIC LEVEL STATE DENSITY FOR PARABOSONIC STRINGS}

\vspace{4mm}

\renewcommand
\baselinestretch{0.8}

{\sc A.A. Bytsenko} \\ {\it Department of Theoretical Physics,
State Technical University, \\ St Petersburg 195251, Russia} \\
{\sc S. D. Odintsov}\footnote{E-mail address: odintsov@ebubecm1.bitnet}
\\
{\it Department E.C.M., Faculty of Physics, University of
Barcelona, \\
Diagonal 647, 08028 Barcelona, Spain} \\
{\sc S. Zerbini} \footnote{E-mail address: zerbini @ itncisca}\\
{\it Department of Physics, University of Trento, 38050 Povo,
Italy}    \\ and
{\it I.N.F.N., Gruppo Collegato di Trento}


\renewcommand
\baselinestretch{1.4}

\vspace{5mm}

\end{center}
\begin{abstract}
Making use of some results concerning the theory of partitions,
relevant in number theory, the complete asymptotic behavior, for
large $n$, of the
level density of states for a parabosonic string is derived.
It is also pointed out the similarity between parabosonic strings and
membranes.
\end{abstract}

\vspace{8mm}


\section{Introduction}

It is quite standard nowdays to describe the quantum field theory in
terms of operators obeying canonical commutation relations. However,
there exists the alternative logical possibility of para-quantum field theory
\cite{gree53-90-270,ohnu82b}, where parafields satisfy tri-linear
commutation relations. Later, the Green's proposal was  investigated in
ref. \cite{gree65-138-1155}. We also would like to remind that parastatistics
is one of the possibilities found by Haag and coworkers
\cite{dopl71-23-199,dopl74-35-49} in a general study of particle
statistics
within the algebraic approach to quantum field theory. Despite the efforts to
apply parastatistics for
the description of internal symmetries (for example, in paraquark
models \cite{ohnu82b}) or even in solid state physics for the
description of quasiparticles, no experimental evidence in favour of
the existence of parafields has been found so far.

Nevertheless, parasymmetry   still can of be of some interest from the
mathematical point of view. For example, it can be
considered as formal extension of the supersymmetry algebra.
Furthermore, some connections with W-symmetry can also be found.
Moreover,  parasymmetry  may find some physical application in string
theory, where parastrings \cite{arda74-9-3341} have been constructed.
It has been showed there that these parastrings possess some
intersesting properties, like the existence of critical dimensions
different from the standard ones, i.e. $D=10$ and $26$.

The present work is devoted to the evaluation of the asymptotic
behaviour of the level state density for  parabosonic strings.
 A complete and  rigorous mathematical result
is obtained,  on the
basis of Meinardus's theorem. Connections with membranes and some
applications are briefly discussed.

\section{Partition Function for Parabosonic Strings}

Here, we will briefly review the paraquantization for parabose harmonic
oscillators relevant to
the parabosonic string, in the limit $p = \infty$, where $p$ is the order of
the paraquantization. The Hamiltonian and the zero point energy for
the free parabose system has the form

\beqn
H& =& \sum_n \frac{\omega_n }{2} (a_n^{\dag} a_n + a_n a_n^{\dag}) - E_0\nn \\
E_0 &=& \frac{p}{2} \sum_n \omega_n\, .
\label{1}
\eeqn
The operators $a_n$ and $a^{\dag} _n$ obey the following tri-commutation
relations \cite{gree53-90-270,ohnu82b}

\beqn
\aq a_n, \ag a^{\dag}_m , a_l \cg \cq &=& 2 \de _{nm} a_l \,,\\
\label{3}
\aq a_n \ag a_m, a_l \cg \cq &=& 0\,.
\label{4}
\eeqn
The vacuum will be chosen to satisfy the relations

\beq
a_n | 0 > = 0, \hs  \ag a^{\dag}_n , a_m \cg \vert 0 > =p \de_{nm} | 0 >\,,
\label{5}
\eeq
so that $H |0 > = 0$. The paracreation operators $a_n^{\dag ,}s$ do
not commute and therefore the Fock space is complicated \cite{ohnu82b}.
For the D-dimensional harmonic oscillators $a_n^i$ of parabosonic
string with frequencies $\omega_n^i = n$, Eqs. (\ref{1}) leads to the
Hamiltonian

\beq
H = \sum^D_{i = 1} \sum^\infty_{n = 1} \frac{n}{2} \ag a^{i \dag}_n,
a^i_n \cg - E_0\,.
\label{(6)}
\eeq
A closed form for the partition function $ Z (t) = \Tr e^{-
tH}$, the trace being
computed over the entire Fock space, in the limit $p \to \ii$, reads
(see, for example, \cite{hama92r})

\begin{equation}
Z(t)=\Tr e^{-tH}=\ag \prod_{ n =1}^\ii \frac{1}{(1-e^{-t n})} \cg ^{D}
\ag \prod_{ n,m =1}^\ii\frac{1} {1-e^{-t (n+m)}}\cg ^{\fr{D^2}{2}}
\ag\prod_{ n =1}^\ii(1-e^{-t2 n})\cg ^{\fr{D}{2}}\,.
\label{7}
\end{equation}

\section{Asymptotic Behavior of the Partition Function}

Our aim is to evaluate, asymptotically, the degeneracy or state level density
corresponding to a parabosonic string, in the limit of infinite
paraquantization order parameter.
As a preliminary result we need the asymptotic expansion of the partition
function for $t \to 0$. To this aim, it may be convenient to work with the
quantity
\beq
F(t)=\log Z(t)=-DF_1(t) +\frac{D}{2} F_1(2t)-\frac{D^2}{2}F_2(t)\,,
\label{8}
\eeq
where we have introduced the definitions
\beqn
F_1(t)&=&\sum_{n=1}^\ii \log(1-e^{-t n})\\
F_2(t)&=&
\sum_{n,m=1}^\ii \log(1-e^{-t( n+m)})\,.
\label{9}
\eeqn
With regards to the two first contributions, one may use the following
result, known in the theory of elliptic modular function
(Hardy-Ramanujan)
\beq
F_1(t)=-\frac{\pi^2}{6t}-
\frac{1}{2}\log (\fr{t}{2\pi})+\frac{t}{24}
+F_1(\fr{4\pi^2}{t})\,.
\label{hr}
\eeq
A simple proof of the above identity is presented in Appendix.

Let us now consider the quantity $F_2(t)$. A  Mellin representation gives

\beq
 \log(1-e^{-t a})=-\frac{1}{2\pi i}
\int_{\Re z=c>2}
dz \Ga(z)\ze(1+z) a^{-z}t^{-z}\, .
\eeq
As a result,
\beq
F_2(t)\equiv \sum_{n,m=1}^\ii \log(1-e^{-t( n+m})=-\frac{1}{2\pi i}
\int_{\Re z=c>2}
dz \Ga(z)\ze(1+z)\ze_2(z)t^{-z}\, ,
\label{91}
\eeq
where
\beq
\ze_2(z)\equiv \sum_{n,m=1}^\ii (n+m)^{-z}\,,
\label{10}
\eeq
$\ze(z)$ being the Riemann zeta function.
Now it is easy to show that (see for example \cite{acto87-20-927})
\beq
\ze_2(z)=\ze(z-1)-\ze(z)\,.
\eeq
So we have
\beq
F_2(t)=G_2(t)-F_1(t)\, ,
\label{11}
\eeq
where
\beq
G_2(t)=-\frac{1}{2\pi i}
\int_{\Re z=c>2}
dz \Ga(z)\ze(1+z)\ze(z-1)t^{-z}\,.
\label{111}
\eeq
Thus, we can deal only with $ G_2(t)$. Concerning this quantity, we
note that one can present it in the form

\beq
G_2(t) =\sum_{n = 1}^\infty \log \at 1 - e^{- tn}\ct^n\,.
\label{12}
\eeq
The related generating function is
\beq
g_2(t) = \prod_{n = 1}^\infty (1 - e^{- tn})^{n}\,.
\label{13}
\eeq
For the estimation of the small $t$ behavior, we can use the
following result on the asymptotic of the partition functions
which admit infinite product as  associated generating function.
To be specific, we shall employ results due to
Meinardus \cite{mein54-59-338,mein54-61-289}.

Let us introduce the generating function
\beq
f(z)=\prod_{ n=1}^\ii[1-e^{-zn}]^{-a_n}\,,
\label{3gen}
\eeq
where $\Re z>0$ and $a_n$ are non-negative real numbers. Let us
consider the associated Dirichlet series
\beq
D(s)=\sum_{n=1}^\ii a_n n^{-s} \, ,\hspace{1cm} \hspace{1cm} s=\si+ir,
\label{diric}
\eeq
which converges for $0<\al <\si$. We assume that $D(s)$ can be
analytically continued in the region $\si \geq -C_o $  ($0<C_o<1$) and
here $D(s)$ is analytic except for a pole of order one at $s=\al$ with residue
$A$.
Besides we assume that
\beq
D(s)=O(|t|^{C_1}),
\eeq
uniformly in $\si \geq -C_o$ as $|t| \rightarrow \ii$, where $C_1$ is
a fixed positive real number.
The following lemma \cite{mein54-59-338,mein54-61-289} is useful with regard
to the asymptotic
properties of $f(z)$, for $z \to 0$, $z=t+2\pi i x$.

{\bf Lemma}

If $f(z)$ and $D(s)$ satisfy the above assumptions, then
\begin{equation}
f(z)=\exp{\{A\Gamma(\a)\zeta(1+\al)z^{-\al}-D(0)\log
z+D'(0)+O(t^{C_o})\}}
\label{l9}
\end{equation}
uniformly in $x$ as $ t \rightarrow 0$, provided $|\arg z|\leq \pi /4$ and
$|x| \leq 1/2$; there exists a positive $\ep$ such that

\begin{equation}
f(t+2\pi i
x)=O(\exp{\{A\Gamma(\al)\zeta(1+\al)t^{-\al}-Ct^{-\varepsilon}\}}),
\label{l10}
\end{equation}
uniformly in $x$ with $y^{\rho} \leq |x| \leq 1/2 $, as $y \rightarrow
0$,  where
\beq
\rho=1+p/2-p\nu/4, \hspace{0.1cm} 0<\nu<2/3,
\eeq
and $C$ a fixed  real number.

We sketch the proof of the  Lemma. One has to use the Mellin-Barnes
representation of the function $ \log f(z)$, namely
\beq
\log f(z)=\frac{1}{2\pi i}\int_{1+\al-i \ii}^{1+\al+i \ii} z^{-s}
\ze(1+s)\Ga(s)D(s)ds\,.
\label{6.15}
\eeq
The integrand in the above equation has a first order pole at $s=\al$ and a
second order pole at $s=0$.  Therefore shifting the vertical contour from
$\Re z=1+\al$ to $\Re z=-C_o$ (due to the conditions of the Lemma the
shift of the line of integration  is permissible) and making use of the
theorem of residues, one obtains
\beqn
\log f(z)&=&A\Gamma(\al)\zeta(1+\al)z^{-\al}-D(0)\log z+D'(0)\nn\\
&+&\frac{1}{2\pi i}\int_{-C_o-i \ii}^{-C_o+i \ii} z^{-s}
\ze(s+1)\Ga(s)D(s)ds \,.
\label{6.16}
\eeqn
The first part of the Lemma follows from  Eq.~
(\ref{6.16}), since the absolute value of the integral in the above
equation can be estimated to behave as $ O(y^{C_o})$. In a
similar way, one can prove the second part of the Lemma
but we do not dwell on this derivation, and we refere to ref.
\cite{andr76b} for details.

Now let us apply the lemma to the
generating function $g_2(t)$. Obviuosly we have $D(s)=-\ze(s-1)$,
$\al=2$, $A=-1$.
According to Meinardus' lemma, for small $t$, we arrive at the following
asymptotic expansion
\beq
G_2(t)(t)\simeq -\ze(3)t^{-2}-\frac{1}{12}\log t -\ze'(-1)+ O(t)\,.
\label{l13}
\eeq
Collecting all the results, we have proved

{ \bf Proposition}
The quantity $F(t)=\log Z(t)$ admits the following asymptotic
expansion, for $t \to 0$
\beqn
F(t)&=&(\frac{D^2}{2}-D)F_1(t)+\frac{D}{2}F_1(2t)
-\frac{D^2}{2}G_2(t)\nn\\
&\simeq & \frac{D^2}{2}\ze(3)t^{-2}+ \log
(t^{\fr{6D-5D^2}{24}}\pi^{\fr{D}{4}}(2\pi)^{\fr{D^2-2D}{4}})+
\frac{D^2}{2}\ze'(-1)+
  O(t^{-1})\,.
\label{P}
\eeqn
As a consequence, the asymptotic behaviour for small $t$ of
the quantity $Z(t)$ reads
\beq
Z(t)\simeq At^B \exp {(Ct^{-2})}\, ,
\label{14}
\eeq
where
\beqn
A&=&\pi^{\fr{D}{4}}(2\pi)^{\fr{D^2-2D}{4}}
e^{\frac{D^2}{2}\ze'(-1)}\nn\\
B&=&\fr{6D-5D^2}{24}\nn\\
C&=&\frac{D^2}{2}\ze(3)\,.
\label{abc}
\eeqn
Note that, in ordinary string theory, the asymptotic behaviour of
$Z_1(t)$ is of the kind $ \exp (ct^{-1})$. We shall see in the next
section the consequence of this different asymptotic behaviour.

\section{Asymptotic Level State Density for Parabosonic String}

The degeneracy or density of levels can easily be calculated starting
from the above asymptotic behaviour.
In fact the density of level for parabosonic string (for a general
discussion on parastrings, see \cite{arda74-9-3341}) may be defined by
\beq
\Tr e^{-zH}=1+\sum_{n=1}^\ii \hat{v}(n)e^{-zn},
\label{15}
\eeq
The Cauchy integral theorem gives
\beq
\hat{v}_n=\frac{1}{2\pi i}\oint e^{zn}Z(z)\,dz \, ,
\label{16}
\eeq
where the contour integral is a small circle about the origin. For $n$ very
large, the leading contribution comes from the asymptotic
behavior for $z \to 0$ of $Z(z)$. Thus,
making use of the Eq. (\ref{14}), we may write
\beq
\hat{v}_n\simeq \frac{A}{2\pi i}\oint z^B e^{zn+Cz^{-2}}\,dz\, .
\label{166}
\eeq
A standard saddle point evaluation, or Meinardus's main theorem, gives
as $n \to \ii$
\beq
\hat{v}_n \simeq \hat{C}_1 n^{-\fr{B+2}{3}} \exp {(\hat{b}_1 n^{\fr{2}{3}})}
\label{17}
\eeq
with
\beq
\hat{C}_1 = A\frac{(2C)^{\fr{2B+1}{6}}}{\sqrt {6\pi}}
\label{18}
\eeq
and
\beq
\hat{b}_1= \frac{3}{2}(2C)^{\fr{1}{3}}=\frac{3}{2}(D^2\ze(3))^{\fr{1}{3}}\, .
\label{19}
\eeq

\section{Concluding remarks}

Eqs.  (\ref{14}), (\ref{abc}) and (\ref{17})-(\ref{19}) are the main result of
this paper. The factor
$\hat{b}_1$ is in
agreement with ref. \cite{hama92r}, where, however, the prefactor
$\hat{C}_1$ was missing. Here, with the help of Meinardus's techniques,
we have been able to compute it.

The asymptotic behaviour given by Eq.  (\ref{17}) should be compared with the
one of the ordinary
bosonic string, whose corresponding  partition function reads

\beq
Z_1 (t) = \prod_{n = 1}^\infty (1 - e^{- tn})^{- a}
\label{(21)}
\eeq
with non-negative $a $. For example, for the open bosonic string,
$a=D-2$. Now $ D (s) = a \zeta
(s), \,\,\al = 1$ and Meinardus's theorems lead to the well known
level density asymptotic behaviour for large $n$ (a derivation, on this
line, can be found in ref. \cite{byts93-304-235})

\beq
v (n) \simeq C_1 n^{- (a + 3)/4} \exp {( b_1 \sqrt n )}\,,
\label{22}
\eeq
where the constants  $C_1$ and $b_1$ are given by

\beq
C_1 = (2 \pi)^{- \fr{a - 1}{2}} (\frac{a}{48})^{1/2}
\label{23}
\eeq
\beq
b_1=\pi \sqrt {\frac{2a}{3}}
\eeq
In this case, since $n\simeq m^2$, the density of states as a function
of mass is (see for example \cite{witt87})
\beq
\rho(m) \simeq m^{\fr{a+1}{2}}\exp {b_1 m}
\eeq
It is well known that this behavior leads to the existence of the
critical Hagedorn temperature, related to the coefficient $b_1$.
For example, the inverse of the Hagedorn temperature for the critical bosonic
string ($D=26$)  is $\be_c=b_1=4\pi$.

On the contrary, if we consider q-branes, higher-dimensional
generalizations
of strings, compactified on a manifold with the topology $(S^1)^q
\times R^{D-q}$, within the semiclassical quantization, one can show
that, asymptotically, the level density behaves
\beq
q(n)\simeq C_q (n)\exp{ B_q n^{\fr{q}{q+1}}}\, ,
\eeq
where the the prefactor $ C_q(n)$ and the coefficient $B_q$ can be
found in ref.  \cite{byts93-304-235}.
As a consequence, the parabosonic string, in the limit of infinite
paraquantization parameter, behaves as an ordinary membrane
(q=2)! In this sense, the parabosonic string seems to have similarity
with black holes, since the asymptotic behavior of the level density
as a function of mass for black holes is similar to the one of q-branes.
There is some indication that canonical partition function for
q-branes does not exist. Thus, with regard to  Hagedorn
temperature, the situation for parabosonic strings
 may be similar to membranes. Hence, the concept of Hagedorn
temperature might be meaningless for parastrings. To answer this
question, one needs to do
careful considerations within the formalism for extended objects at non-zero T,
 which has not developed yet.

\ack{ SDO thanks the Generalitat de Catalonya for financial support }

\section{Appendix}

The proof of the Hardy-Ramanujan formula can be done by making use of the
Mellin representation, namely
\beq
F_1(t)\equiv \sum_{n=1}^\ii \log(1-e^{-t n})=-\frac{1}{2\pi i}\int_{\Re z=c}
dz \Ga(z)\ze(1+z)\ze(z)t^{-z}\, ,
\label{6}
\eeq
with $c>1 $, and $\zeta (s)$ is the Riemann zeta-function. Shifting the line
of integration from $\Re z=c>1$ to $ \Re z =c'$,
 $ -1 <c'<0$, noting that the integrand has a first-order pole at $z=1$
and a second-order pole at $ z=0$ one arrives at
\beq
F_1(t)=-\frac{\pi^2}{6t}-
\frac{1}{2}\log (\fr{t}{2\pi})-\frac{1}{2\pi i}\int_{\Re z=c'}
dz \Ga(z)\ze(1+z)\ze(z)t^{-z}\, .
\label{77}
\eeq
Making the change of variable in the complex integral $ z=-s$ and
using the functional equations for the functions $\Ga(s)$ and $\ze(s)$
we get
\beq
F_1(t)=-\frac{\pi^2}{6t}-
\frac{1}{2}\log (\fr{t}{2\pi})-\frac{1}{2\pi i}\int_{\Re z=c''}
dz \Ga(z)\ze(1+z)\ze(z)(\fr{4\pi^2}{t})^{-z}\, .
\label{88}
\eeq
where $ 0<c''<1$. The identity (\ref{5}) is obtained shifting the line
of integration to $\Re z=c'''>1$, and taking into account of the
first-order pole at $z=1$.

\end{document}